\documentclass[preprint,12pt]{elsarticle}
\usepackage{lineno}
\modulolinenumbers[5]

\usepackage{geometry}                
\usepackage{subfig}
\usepackage{placeins}
\usepackage{amssymb}

\usepackage[multidot]{grffile}

\journal{Nuclear Instruments and Methods A}

\begin{document}
                  
\begin{frontmatter}

\title{A Poisson likelihood approach to fake lepton estimation with the matrix method}

\author{Erich W. Varnes\corref{}}
\ead{varnes@physics.arizona.edu}
\address{University of Arizona, 1118 E. 4th St., Tucson AZ 85721, USA}

\begin{abstract}
Many high-energy physics analyses require the presence of leptons from $W$, $Z$, or $H$ boson decay. For these analyses, signatures that mimic such 
leptons present a `fake lepton' background that must be estimated.  Since the magnitude of this background depends strongly upon details of the detector 
response, it can be difficult to estimate with simulation.  One data-driven approach is the `matrix method', in which two categories of leptons are defined (`loose' and `tight'), with the tight category being a subset of the loose category.  Using the populations of leptons in each category in the analysis sample, and the efficiencies for both real and fake leptons in the loose category to satisfy the criteria for the tight category, the fake background yield can be estimated.  This paper describes a Poisson likelihood implementation of the matrix method, which provides a more precise, reliable, and robust estimate of the fake background yield compared to an analytic solution.  This implementation also provides a reliable estimate of the background for cases in which the analysis selection permits more loose leptons than tight leptons, potentially allowing for greater selection efficiency.
\end{abstract}

\end{frontmatter}


\section{Introduction}

Many experimental particle physics analyses depend upon the identification of leptons\footnote{`Lepton' is taken here to mean electron or muon}, typically those from the decay of a $W$ or $Z$ boson.  Such analyses are in general complicated by the fact that other processes, involving for example leptons from the decay of heavy quarks or, in the case of electrons, photon conversions or hadronic jets with a large electromagnetic energy fraction, can give rise to detector signatures that are difficult to distinguish from those of the desired leptons (for simplicity, the desired leptons will hereafter be called `real leptons' and the mimicking signatures `fake leptons').  The magnitude and properties of the fake lepton background are difficult to estimate with simulation from first principles, and therefore data-driven techniques are often employed.  One of these techniques, known as the `matrix method' and first described in Ref.~\cite{Abbott:1999tt}, depends on employing two levels of lepton identification criteria.  One of these, called the `tight' criteria, are simply those that are used to identify leptons in the analysis (e.g. in the sample for which the fake lepton background is to be determined), while the other is a less restrictive set, called the `loose' criteria, defined so that every event selected with the tight criteria will also be selected with the loose criteria.  If the efficiencies for both fake and real leptons that satisfy the loose criteria to also satisfy the tight criteria are known, the number of fake lepton events in the tight sample can be deduced from the numbers of loose and tight events. This paper presents a new method for performing this deduction that offers better precision, more accurate uncertainties, and is more robust than methods that are typically currently employed.

\section{The Matrix Method}

For an analysis of dilepton events\footnote{The matrix method can in principle be applied to events with any number of leptons; dilepton events will be used in most examples here since they provide non-trivial complexity while still permitting the relevant equations to be reasonably compact.}, the experimental quantities are the efficiencies $r$ and $f$ for real and fake leptons, respectively, to satisfy the tight selection criteria, and the whether or not each lepton in the sample satisfies the tight criteria.  Using the notation $\bar{t}$ to represent leptons that do not satisfy the tight criteria, the known quantities are related to the number of events in the tight sample with fake leptons ($N^{tt}_{\mathrm {fake}}$) by: 
\begin{equation}
\left( {\begin{array}{*{20}{c}}
{{N^{tt}}}\\
{{N^{t\bar t}}}\\
{{N^{\bar tt}}}\\
{{N^{\bar t\bar t}}}
\end{array}} \right) = \left( {\begin{array}{*{20}{c}}
{{r_1}{r_2}}&{{r_1}{f_2}}&{{f_1}{r_2}}&{{f_1}{f_2}}\\
{{r_1}{{\tilde r}_2}}&{{r_1}{{\tilde f}_2}}&{{f_1}{{\tilde r}_2}}&{{f_1}{{\tilde f}_2}}\\
{{{\tilde r}_1}{r_2}}&{{{\tilde r}_1}{f_2}}&{{{\tilde f}_1}{r_2}}&{{{\tilde f}_1}{f_2}}\\
{{{\tilde r}_1}{{\tilde r}_2}}&{{{\tilde r}_1}{{\tilde f}_2}}&{{{\tilde f}_1}{{\tilde r}_2}}&{{{\tilde f}_1}{{\tilde f}_2}}
\end{array}} \right)\left( {\begin{array}{*{20}{c}}
{N_{rr}^{tt}/\left( {{r_1}{r_2}} \right)}\\
{N_{rf}^{tt}/\left( {{r_1}{f_2}} \right)}\\
{N_{fr}^{tt}/\left( {{f_1}{r_2}} \right)}\\
{N_{ff}^{tt}/\left( {{f_1}{r_2}} \right)}
\end{array}} \right)
\end{equation}
where $\tilde{r} = 1-r$ and $\tilde{f} = 1-f$, and $N^{tt}_{\mathrm {fake}} = N_{rf}^{tt} + N_{fr}^{tt} + N_{ff}^{tt}$.  The indices on $r$ and $f$ reflect the fact that the efficiencies can vary significantly depending on features of the leptons or of the events in which they appear, and therefore $r$ and $f$ are estimated separately for each lepton in the event (the indices 1 and 2 typically refer to the $p_T$ rank of the lepton).  Solving the above system of equations for $N^{tt}_\mathrm{fake}$ gives:

\begin{eqnarray}
N^\mathrm{t}_{\mathrm{fake},i}  & = & \alpha r_{1,i} f_{2,i} \left[-\tilde{f_{1,i}}\tilde{r_{2,i}}N^\mathrm{tt} + \tilde{f_{1,i}}r_{2,i}N^\mathrm{t\overline{t}}+f_{1,i}\tilde{r_{2,i}}N^\mathrm{\overline{t}t}-f_{1,i}r_{2,i}N^\mathrm{\overline{t}\overline{t}}\right] \nonumber \\
& & + \alpha f_{1,i} r_{2,i} \left[-\tilde{r_{1,i}}\tilde{f_{2,i}}N^\mathrm{tt} + \tilde{r_{1,i}}f_{2,i}N^\mathrm{t\overline{t}} + r_{1,i}\tilde{f_{2,i}}N^\mathrm{\overline{t}t} - r_{1,i}f_{2,i}N^\mathrm{\overline{t}\overline{t}}\right] \nonumber \\
& & + \alpha f_{1,i} f_{2,i} \left[\tilde{r_{1,i}}\tilde{r_{2,i}}N^\mathrm{tt} - \tilde{r_{1,i}}r_{2,i}N^\mathrm{t\overline{t}}-r_{1,i}\tilde{r_{2,i}}N^\mathrm{\overline{t}t}+r_{1,i}r_{2,i}N^\mathrm{\overline{t}\overline{t}}\right] \nonumber
\label{eqn:intro-mm-dilepton-fake}
\end{eqnarray}
where
\[\alpha = \frac{1}{(r_{1,i} - f_{1,i})(r_{2,i} - f_{2,i})} .\]
\\
In standard implementations of the matrix method, the above quantity is generally calculated for each event (where one of $N^\mathrm{tt}$, $N^\mathrm{t\overline{t}}$, $N^\mathrm{\overline{t}t}$, and $ N^\mathrm{\overline{t}\overline{t}}$ is one and the others are zero).   Then the 
number of fake lepton events in the sample and its uncertainty are estimated as:

\begin{equation}
\label{eqn:stdMM} 
N^\mathrm{t}_{\mathrm{fake}} = \sum{N^\mathrm{t}_{\mathrm{fake},i}} \pm \sqrt{ \sum{(N^\mathrm{t}_{\mathrm{fake},i}})^2}.
\end{equation}

The standard matrix method has three shortcomings, which are most apparent for low-statistics samples: $i$) the estimated central value for the number of fake lepton events can be negative, which causes difficulty in interpretation,  $ii$) it can be numerically unstable if $r$ and $f$ have similar values for any of the leptons in the sample (due to the $r-f$ terms that appear in the denominator of $\alpha$), and $iii$) the uncertainty calculated via Eq.~\ref{eqn:stdMM} can be unreliable.  Issue $ii$) can be mitigated by taking the average of the products of $r$ and $f$ that appear in~$\ref{eqn:intro-mm-dilepton-fake}$ for the entire sample rather than calculating $N^\mathrm{t}_{\mathrm{fake},i}$ for each event.  The Poisson likelihood approach here will be compared to both of the above variants of the matrix method, which are hereafter referred to as the `standard matrix method' and the `standard matrix method with average efficiencies'.

\section{The Likelihood Matrix Method}

All of the shortcomings discussed above can be addressed by the use of a maximum likelihood approach. In general, the likelihood can include both Gaussian constraints on the values of $r$ and $f$, and Poisson constraints on $N^{tt}$, $N^{t\bar{t}}$, $N^{\bar{t}t}$, and $N^{tt}$:

\begin{equation}
\label{eq:fullL}
L = G(r, \sigma_r)G(f, \sigma_f)P(N^{tt}, N^{tt}_{\mathrm {pred}})P(N^{t\bar{t}}, N^{t\bar{t}}_{\mathrm {pred}})P(N^{\bar{t}t}, N^{\bar{t}t}_{\mathrm {pred}})P(N^{\bar{t}\bar{t}}, N^{\bar{t}\bar{t}}_{\mathrm {pred}})
\end{equation}
where $G$ represents Gaussian constraints on $r$ and $f$ given their uncertainties $\sigma_r$ and $\sigma_f$, and $P$ represents Poisson constraints on the observed numbers of events in each lepton quality category be consistent with the presumed numbers of real and fake leptons \cite{read_lewis}.  While this equation is complete, its implementation suffers from the fact that $r$ and $f$ will in general be different for every lepton in the sample, meaning that the number of parameters to be fit can become large, leading to unstable and time-consuming fits.  One possible procedure that has been explored is to define a small number of ``categories'' of leptons with similar $r$ and $f$ values~\cite{Gillam:2014xua}.  In this paper a different approach is pursued, in which the uncertainties on $r$ and $f$ are ignored, and only the Poisson terms in the likelihood are considered, so that Eq.~\ref{eq:fullL} is reduced to
\begin{equation}
L = P(N^{tt}, N^{tt}_{\mathrm {pred}})P(N^{t\bar{t}}, N^{t\bar{t}}_{\mathrm {pred}})P(N^{\bar{t}t}, N^{\bar{t}t}_{\mathrm {pred}})P(N^{\bar{t}\bar{t}}, N^{\bar{t}\bar{t}}_{\mathrm {pred}}).
\end{equation}
  To evaluate those terms, the following relationship between the numbers of real and fake leptons and the predicted numbers of events in each lepton quality category is used:
\begin{equation}
\label{eq:lhoodMM_nfake_loose}
\begin{array}{l}
N_{{\mathrm{pred}}}^{tt}  = \langle r_1r_2 \rangle N_{rr}^{ll} + \langle r_1f_2 \rangle N_{rf}^{ll} + \langle f_1r_2 \rangle N_{fr}^{ll} + \langle f_1f_2 \rangle N_{ff}^{ll}  \\
N_{{\mathrm{pred}}}^{t\bar{t}}  = \langle r_1\tilde{r_2} \rangle N_{rr}^{ll} + \langle r_1 \tilde{f_2} \rangle N_{rf}^{ll} + \langle f_1 \tilde{r_2} \rangle N_{fr}^{l} + \langle f_1\tilde{f_2} \rangle N_{ff}^{ll}  \\
N_{{\mathrm{pred}}}^{\bar tt} = \langle \tilde{r_1}r_2 \rangle N_{rr}^{ll} + \langle \tilde{ r_1} f_2 \rangle N_{rf}^{ll} + \langle \tilde{f_1}r_2 \rangle N_{fr}^{ll} + \langle \tilde{f_1}f_2 \rangle N_{ff}^{ll}  \\
N_{{\mathrm{pred}}}^{\bar t\bar t} = \langle \tilde{r_1}\tilde{r_2} \rangle N_{rr}^{ll} + \langle \tilde{r_1}\tilde{f_2} \rangle N_{rf}^{ll} + \langle \tilde{f_1}\tilde{r_2} \rangle N_{fr}^{ll} + \langle \tilde{f_1}\tilde{f_2} \rangle N_{ff}^{ll}  \\
\end{array}
\end{equation}
where $\left\langle {{\epsilon_1}{\epsilon_2}} \right\rangle$ represents the average  of the product of the relevant quantities in the loose lepton sample.  While the above is expressed in terms of the real and fake lepton contributions to the loosely-selected sample, the desired quantities are the contributions to the tightly-selected sample, so the above relations are recast as:

\begin{equation}
\label{eq:lhoodMM_nfake_tight}
\begin{array}{l}
N_{{\mathrm{pred}}}^{tt}  =  N_{rr}^{tt} +  N_{rf}^{tt} +  N_{fr}^{tt} +  N_{ff}^{tt}  \\
N_{{\mathrm{pred}}}^{t\bar{t}}  = {\langle r_1\tilde{r_2} \rangle \over \langle r_1r_2 \rangle} N_{rr}^{tt} +{ \langle r_1 \tilde{f_2} \rangle \over \langle r_1f_2 \rangle} N_{rf}^{tt} + {\langle f_1 \tilde{r_2} \rangle  \over  \langle f_1r_2 \rangle } N_{fr}^{tt} + { \langle f_1\tilde{f_2} \rangle \over \langle f_1f_2 \rangle }N_{ff}^{tt}  \\
N_{{\mathrm{pred}}}^{\bar tt} ={ \langle \tilde{r_1}r_2 \rangle \over \langle r_1r_2 \rangle} N_{rr}^{tt} + { \langle \tilde{ r_1} f_2 \rangle \over \langle r_1f_2 \rangle } N_{rf}^{tt} + { \langle \tilde{f_1}r_2 \rangle \over  \langle f_1r_2 \rangle}  N_{fr}^{tt} + { \langle \tilde{f_1}f_2 \rangle \over  \langle f_1f_2 \rangle}N_{ff}^{tt}  \\
N_{{\mathrm{pred}}}^{\bar t\bar t} ={  \langle \tilde{r_1}\tilde{r_2} \rangle \over \langle r_1r_2 \rangle}  N_{rr}^{tt} +{  \langle \tilde{r_1}\tilde{f_2} \rangle \over \langle r_1f_2 \rangle}  N_{rf}^{tt} +{ \langle \tilde{f_1}\tilde{r_2} \rangle \over  \langle f_1r_2 \rangle} N_{fr}^{tt} + { \langle \tilde{f_1}\tilde{f_2} \rangle  \over \langle f_1f_2 \rangle} N_{ff}^{tt}  \\
\end{array}
\end{equation}

With the  numbers of tight and non-tight leptons predicted for given values of $N_{rr}$, $N_{rf}$, $N_{fr}$, and $N_{ff}$, the simplified likelihood
\begin{equation}
\label{eq:lhoodMM_def}
L = P(N^{tt}, N^{tt}_{\mathrm {pred}})P(N^{t\bar{t}}, N^{t\bar{t}}_{\mathrm {pred}})P(N^{\bar{t}t}, N^{\bar{t}t}_{\mathrm {pred}})P(N^{\bar{t}\bar{t}}, N^{\bar{t}\bar{t}}_{\mathrm {pred}})
\end{equation}
can be used.  Once the likelihood is minimized, the total number of events with at least one fake lepton in the sample of events with two tight leptons is given by:

\begin{equation}
N_{\mathrm {fake}}^{tt} = N_{rf}^{tt} +  N_{fr}^{tt} +  N_{ff}^{tt}  
\end{equation}

 The results in this paper were obtained using the {\sc minuit} function minimization package~\cite{James:1975dr}, as implemented via the TMinuit class in {\sc root}~\cite{Brun:1997pa}.

A variation on the method can also be used when the number of tight leptons required by the analysis is less than the number of loose leptons considered.  For example, an analysis may accept events with either one or two tight leptons.  An example might be an analysis of $t\bar{t}$ pair production in the  `dilepton' final state $t\bar{t} \rightarrow W^+bW^-\bar{b} \rightarrow \ell+\nu b \ell^{\prime -}\nu_{\ell^\prime}$, which might permit events with only one tight lepton (to maximize efficiency) but where it would not make sense to reject events with one tight and one loose lepton. In such cases, the contribution to the fake yield from events with two loose leptons includes cases where either one or both of the loose leptons is tight.  This contribution is denoted as $N_{\mathrm {fake}}^{t+tt}$, and is the sum of $N_{rf}^{tt}$, $N_{rf}^{\bar{t}t}$, $N_{fr}^{tt}$, $N_{fr}^{t\bar{t}}$, $N_{rf}^{tt}$, $N_{ff}^{t\bar{t}}$, and $N_{ff}^{\bar{t}t}$.  This means that the natural choice of fit parameters in this case is the set of $N_{rr}^{tt}$ and

\begin{equation}
\label{eq:lhoodMM_nfake_params_1lepincl}
\begin{array}{lcl}
N_{rf}^{t+tt} &\equiv&  N_{rf}^{tt} + N_{rf}^{\bar{t}t} = \left( \langle r_1 f_2 \rangle + \langle \tilde{r_1}f_2 \rangle \right) N_{rf}^{ll} = \langle f_2 \rangle N_{rf}^{ll} = { \langle f_2 \rangle  \over \langle r_1 f_2 \rangle } N_{rf}^{tt} \\
N_{fr}^{t+tt} &\equiv&  N_{fr}^{tt} + N_{fr}^{t\bar{t}} = \left( \langle f_1 r_2 \rangle + \langle f_1\tilde{r_2} \rangle \right) N_{fr}^{ll} = \langle f_1 \rangle N_{fr}^{ll} =  {  \langle f_1\rangle  \over \langle f_1 r_2 \rangle } N_{fr}^{tt} \\
N_{ff}^{t+tt} &\equiv&  N_{ff}^{tt} + N_{ff}^{t\bar{t}} +  N_{ff}^{\bar{t}t} = \left( \langle f_1 f_2 \rangle + \langle f_1\tilde{f_2} \rangle + \langle \tilde{f_1}f_2 \rangle \right) N_{ff}^{ll} \\ 
 &=& \left( \langle f_1\rangle+\langle f_2 \rangle - \langle  f_1f_2 \rangle \right) N_{ff}^{ll}  = {   \langle f_1\rangle+ \langle f_2 \rangle - \langle  f_1f_2 \rangle  \over \langle f_1 f_2 \rangle } N_{ff}^{tt},
\end{array}
\end{equation}

so that $N_{\mathrm {fake}}^{t+tt} = N_{rf}^{t+tt}  + N_{fr}^{t+tt}  + N_{ff}^{t+tt}$  
and the appropriate form of Eq.~\ref{eq:lhoodMM_nfake_tight}  is

\begin{equation}
\label{eq:lhoodMM_nfake_ge1_tight}
\begin{array}{l}
N_{{\mathrm{pred}}}^{tt}  =  N_{rr}^{tt} +  {  \langle r_1 f_2 \rangle \over  \langle f_2 \rangle }N_{rf}^{t+tt} +  {\langle f_1 r_2 \rangle   \over   \langle f_1  \rangle} N_{fr}^{t+tt} +{ \langle f_1 f_2 \rangle \over  \langle f_1\rangle+ \langle f_2 \rangle - \langle  f_1f_2 \rangle  } N_{ff}^{t+tt}  \\
N_{{\mathrm{pred}}}^{t\bar{t}}  = {\langle r_1\tilde{r_2} \rangle \over \langle r_1r_2 \rangle} N_{rr}^{tt} +{ \langle r_1 \tilde{f_2} \rangle \over \langle f_2 \rangle} N_{rf}^{t+tt } + {\langle f_1 \tilde{r_2} \rangle  \over  \langle f_1 \rangle } N_{fr}^{t+tt} + { \langle f_1\tilde{f_2} \rangle \over \langle f_1\rangle + \langle f_2 \rangle - \langle f_1f_2 \rangle }N_{ff}^{t+tt}  \\
N_{{\mathrm{pred}}}^{\bar tt} ={ \langle \tilde{r_1}r_2 \rangle \over \langle r_1r_2 \rangle} N_{rr}^{tt} + { \langle \tilde{ r_1} f_2 \rangle \over \langle f_2 \rangle  } N_{rf}^{t+tt} + { \langle \tilde{f_1}r_2 \rangle \over \langle f_1 \rangle }  N_{fr}^{t+tt} + { \langle \tilde{f_1}f_2 \rangle \over   \langle f_1\rangle + \langle f_2 \rangle - \langle f_1f_2 \rangle }N_{ff}^{t+tt}  \\
N_{{\mathrm{pred}}}^{\bar t\bar t} ={  \langle \tilde{r_1}\tilde{r_2} \rangle \over \langle r_1r_2 \rangle}  N_{rr}^{tt} +{  \langle \tilde{r_1}\tilde{f_2} \rangle \over \langle f_2 \rangle}  N_{rf}^{t+tt} +{ \langle \tilde{f_1}\tilde{r_2} \rangle \over  \langle f_1 \rangle} N_{fr}^{t+tt} + { \langle \tilde{f_1}\tilde{f_2} \rangle  \over \langle f_1\rangle + \langle f_2 \rangle - \langle f_1f_2 \rangle } N_{ff}^{t+tt}.  \\
\end{array}
\end{equation}

Similarly, an analysis may choose to require exactly one tight lepton but permit events that have additional loose leptons.  For the case that one additional loose lepton is permitted, the contribution to the fake yield from events with two loose leptons includes only cases where exactly one of the leptons is tight.  This contribution is denoted as  $N^t_{\mathrm {fake}}$, and is the sum of  $N_{rf}^{t} $, $N_{rf}^{t}$, and  $N_{ff}^{t}$ where

\begin{equation}
\label{eq:lhoodMM_nfake_params_1lepexcl}
\begin{array}{lcl}
N_{rf}^{t} &\equiv&  N_{rf}^{\bar{t}t} =  \langle \tilde{r_1}f_2 \rangle N_{rf}^{ll}  = { \langle \tilde{r_1}f_2 \rangle  \over \langle r_1 f_2 \rangle } N_{rf}^{tt} \\
N_{fr}^{t} &\equiv&  N_{fr}^{t\bar{t}} = \langle f_1\tilde{r_2} \rangle  N_{fr}^{ll} =   {   \langle f_1\tilde{r_2} \rangle \over \langle f_1 r_2 \rangle } N_{fr}^{tt} \\
N_{ff}^{t} &\equiv&  N_{ff}^{t\bar{t}} +  N_{ff}^{\bar{t}t} = \left( \langle f_1\tilde{f_2} \rangle + \langle \tilde{f_1}f_2 \rangle \right) N_{ff}^{ll} \\ 
 &=& \left( \langle f_1\rangle+\langle f_2 \rangle - 2\langle  f_1f_2 \rangle \right) N_{ff}^{ll}  = {   \langle f_1\rangle+ \langle f_2 \rangle - 2 \langle  f_1f_2 \rangle  \over \langle f_1 f_2 \rangle } N_{ff}^{tt},
\end{array}
\end{equation}
and the appropriate form of Eq.~\ref{eq:lhoodMM_nfake_tight} becomes 

\begin{equation}
\label{eq:lhoodMM_nfake_eq1_tight}
\begin{array}{l}
N_{{\mathrm{pred}}}^{tt}  =  N_{rr}^{tt} +  {  \langle r_1 f_2 \rangle \over  \langle \tilde{r_1} f_2 \rangle }N_{rf}^{t} +  {\langle f_1 r_2 \rangle   \over   \langle f_1 \tilde{f_2}  \rangle} N_{fr}^{t} +{ \langle f_1 f_2 \rangle \over  \langle f_1\rangle+ \langle f_2 \rangle -2 \langle  f_1f_2 \rangle  } N_{ff}^{t}  \\
N_{{\mathrm{pred}}}^{t\bar{t}}  = {\langle r_1\tilde{r_2} \rangle \over \langle r_1r_2 \rangle} N_{rr}^{tt} +{ \langle r_1 \tilde{f_2} \rangle \over \langle \tilde{r_1}f_2 \rangle} N_{rf}^{t } + {\langle f_1 \tilde{r_2} \rangle  \over  \langle f_1 \tilde{r_2} \rangle } N_{fr}^{t} + { \langle f_1\tilde{f_2} \rangle \over \langle f_1\rangle + \langle f_2 \rangle -2 \langle f_1f_2 \rangle }N_{ff}^{t}  \\
N_{{\mathrm{pred}}}^{\bar tt} ={ \langle \tilde{r_1}r_2 \rangle \over \langle r_1r_2 \rangle} N_{rr}^{tt} + { \langle \tilde{ r_1} f_2 \rangle \over \langle \tilde{r_1} f_2 \rangle  } N_{rf}^{t} + { \langle \tilde{f_1}r_2 \rangle \over \langle f_1\tilde{r_2} \rangle }  N_{fr}^{t} + { \langle \tilde{f_1}f_2 \rangle \over   \langle f_1\rangle + \langle f_2 \rangle - 2\langle f_1f_2 \rangle }N_{ff}^{t}  \\
N_{{\mathrm{pred}}}^{\bar t\bar t} ={  \langle \tilde{r_1}\tilde{r_2} \rangle \over \langle r_1r_2 \rangle}  N_{rr}^{tt} +{  \langle \tilde{r_1}\tilde{f_2} \rangle \over \langle \tilde{r_1}f_2 \rangle}  N_{rf}^{t} +{ \langle \tilde{f_1}\tilde{r_2} \rangle \over  \langle f_1 \tilde{r_2}  \rangle} N_{fr}^{t} + { \langle \tilde{f_1}\tilde{f_2} \rangle  \over \langle f_1 \rangle  + \langle f_2 \rangle - 2\langle f_1f_2 \rangle} N_{ff}^{t}.  \\
\end{array}
\end{equation}
In both of the above cases the contribution from the two loose lepton sample would be added to the contribution from the sample with one loose lepton to find the total fake yield. Therefore the method can be adapted to provide the correct fake background estimate for a variety of requirements on the numbers of tight and loose leptons.

\subsection{Numerical stability}

A potential drawback of a maximum likelihood fit is the possibility of the fit failing to converge, or converging to a local minimum rather than the global minimum.  In addition, solutions where any component that contributes to the fake yield is negative are to be avoided as unphysical.  To facilitate this, the parameters are transformed such that they represent a vector where the magnitude of the vector is $\sqrt{N_{\mathrm {fake}}^{tt}}$ and the cartesian components of the vector are $\sqrt{N_{rf}^{tt}}$, $\sqrt{N_{fr}^{tt}}$, and $\sqrt{N_{ff}^{tt}}$.  The parameters of the fit are then $N_{\mathrm {fake}}^{tt}$, and angles $\theta$ and $\phi$ representing the direction of the vector.  By constraining both $\theta$ and $\phi$ to be between 0 and $\pi/2$ and constraining $N_{\mathrm {fake}}^{tt}$ to be greater than 0 and less than the number of events in the loose sample, the fit is limited to solutions with the desired characteristics.  This approach has the further advantage that the fit directly returns the value of $N^{tt}_{\mathrm {fake}}$ and its uncertainty, rather than requiring the analyzer to derive them from the values and uncertainties of each component.

\subsection{Computing time}
One advantage of the standard matrix method is that it is computationally simple, and therefore can be done very quickly.  The likelihood matrix method is unavoidably slower, since it involves several iterations of evaluating Eqs.~\ref{eq:lhoodMM_nfake_tight} and \ref{eq:lhoodMM_def} as the minimum of $-\ln L$ is found.  However, the evaluation of these functions is fast since the coefficients in Eq.~\ref{eq:lhoodMM_nfake_tight} only need to be evaluated once (this is a consequence of treating the $r_i$ and $f_i$ as fixed quantities rather than as fit parameters).  As a result, fits of dilepton events, with 1000 events in the loose lepton sample, take $\approx 4$ ms per fit on an Intel Xeon processor running at 2.7 GHz.  This means that the likelihood calculation is fast enough that it will not present a practical impediment to most physics analyses, even when many fits must be done to, for example, determine the fake lepton contribution to each of many bins in a distribution.

\section{Comparison of the standard and likelihood-based matrix methods}

\subsection{Toy Monte Carlo}

To assess the performance of the method, simulations of experiments with different sample sizes are prepared (consisting of 5, 100, or 1000 events in the loose lepton sample).  The fraction of fake leptons in the loose lepton sample is varied for each pseudo experiment, with a uniform distribution between 0 and 95\% (samples consisting of more than 95\% fake leptons are unlikely to be of interest in physics analyses).  The values of $r$ and $f$ for each lepton are drawn from Gaussian distributions with means of 0.90 and 0.20, respectively, and widths of 0.10 (limits are placed such that the values are always between 0 and 1, and $f$ is always at least 1\% less than $r$).  Each lepton is randomly assigned as fake or real, according to the fraction of fake leptons assumed for the pseudoexperiement.  Then each lepton is assigned as passing or failing the tight selection criterial based on whether or not it is a real lepton, and the values of $r$ and $f$ assigned to it.  The expectation value for the number of fake lepton events in each sample is then given by:

\begin{equation} 
\langle N^{tt}_{\mathrm {fake}} \rangle = \sum_{rf \hbox{ events}}r_{1,i}f_{2,i} +  \sum_{fr \hbox{ events}}f_{1,i}r_{2,i} +    \sum_{ff \hbox{ events}} f_{1,i}f_{2,i} 
\end{equation}

It is this expectation value that the matrix method is intended to estimate, rather than the actual number of fake leptons in the sample.  The latter is subject to Poisson fluctuations which are typically accounted for when assigning the uncertainties on the physical observable of interest (e.g., a production cross section for a signal).

\subsection{Results}

\begin{figure}
\begin{centering}
  \includegraphics[width=\textwidth]{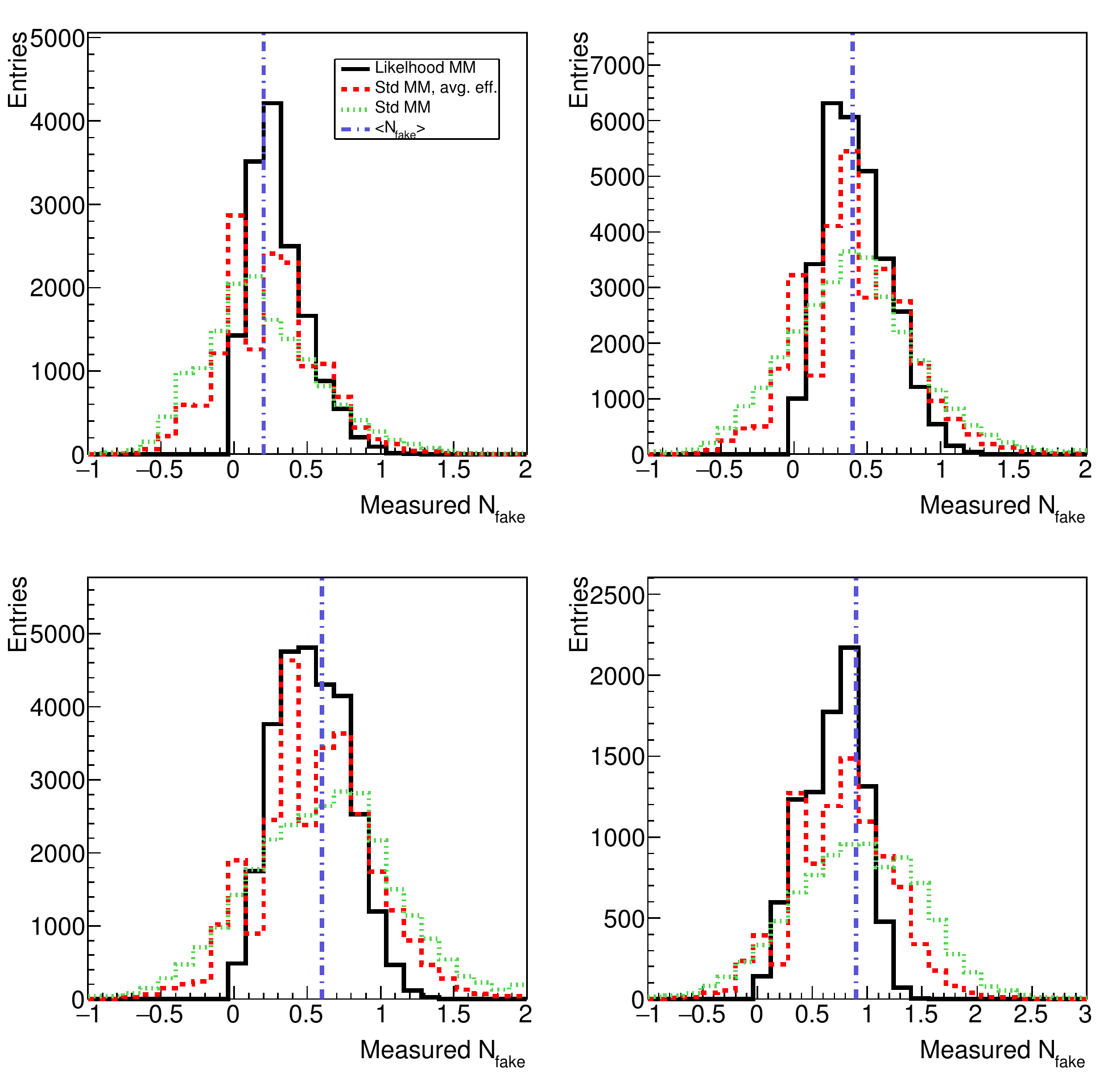}
\end{centering}
\caption{\label{fig:nfake_5ev_lowfe}Distribution of the estimated fake yield in toy MC samples consisting of five events with two loose leptons per event, for the standard matrix method, the standard matrix method with average efficiencies, and the likelihood matrix method.  In these samples, the average value for $r$ is 0.90 and for $f$ is 0.20, with each efficiency having a Gaussian-distributed spread of 0.10.  The four plots show samples with different average values for the true number of events with fake leptons, as indicated by the dashed lines. }
\end{figure}

\begin{figure}
  \begin{center}
  \includegraphics[width=\textwidth]{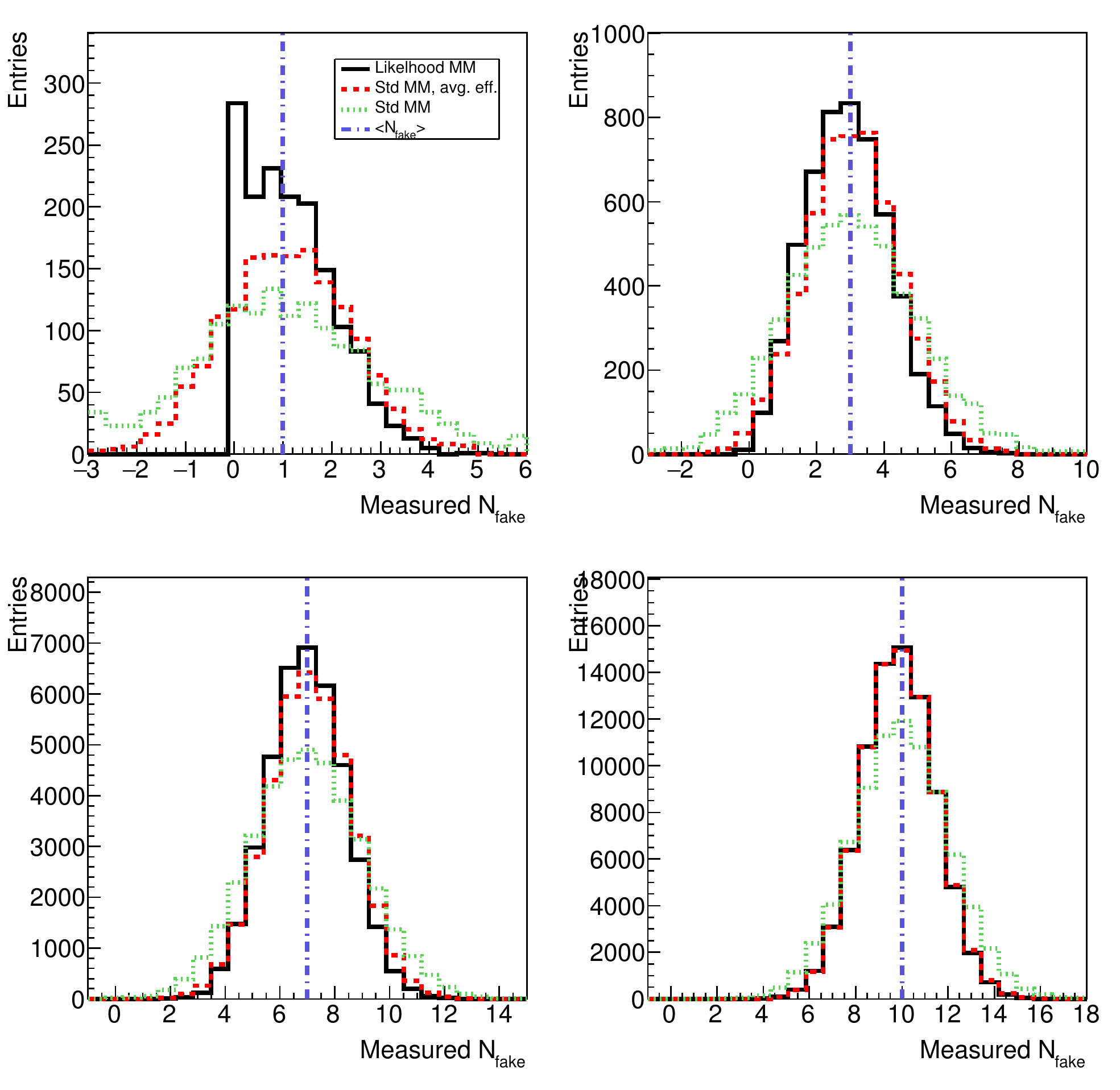}
  \end{center}
  \caption{\label{fig:nfake_100ev_lowfe}Distribution of the estimated fake yield in toy MC samples consisting of 100 events with two loose leptons per event, for  the standard matrix method, the standard matrix method with average efficiencies,  and the likelihood matrix method.  In these samples, the average value for $r$ is 0.90 and for $f$ is 0.20, with each efficiency having a Gaussian-distributed spread of 0.10.  The four plots show samples with different average values for the true number of events with fake leptons, as indicated by the dashed lines. }
\end{figure}

\begin{figure}
  \begin{center}
  \includegraphics[width=\textwidth]{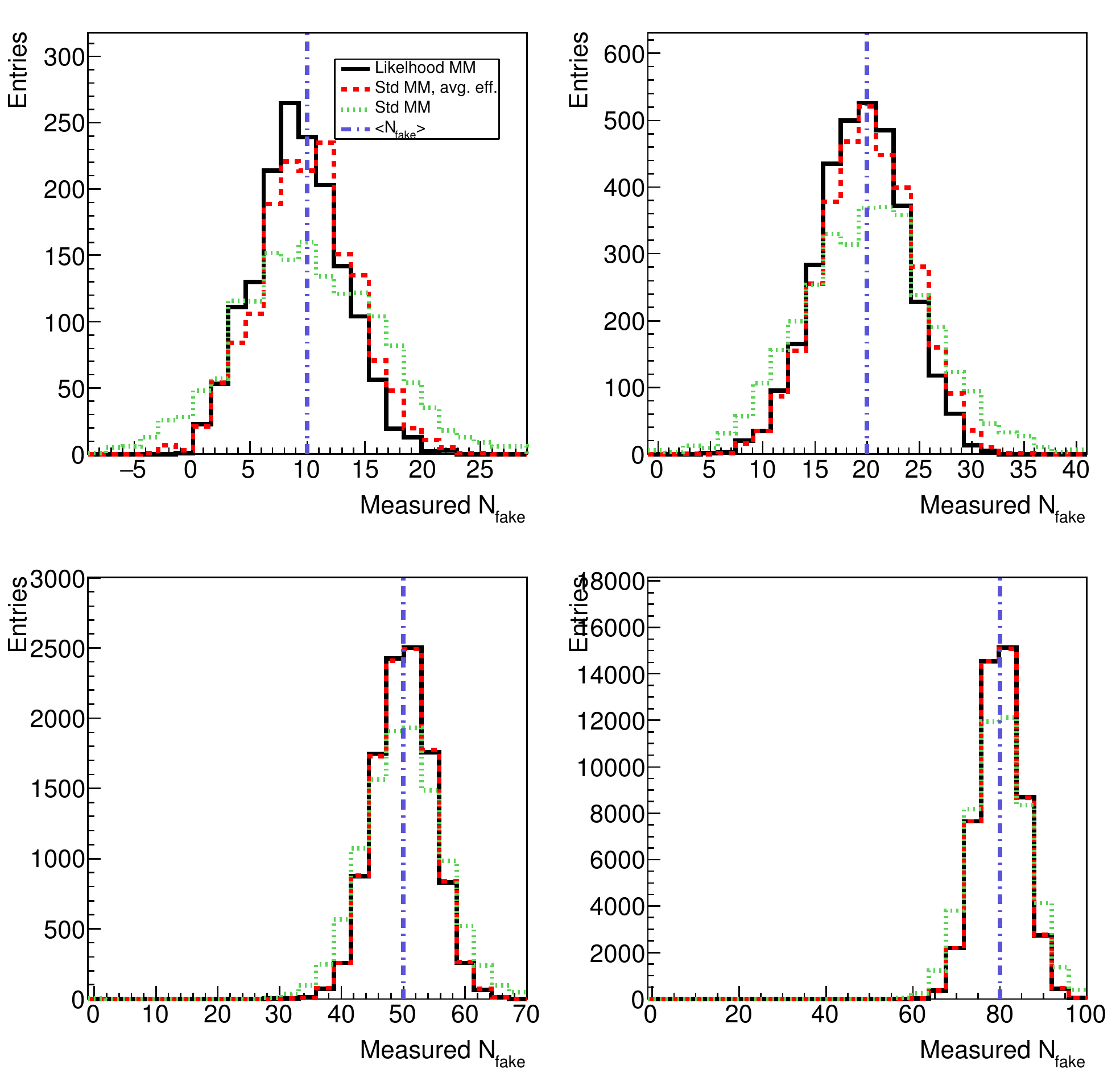}
  \end{center}
  \caption{\label{fig:nfake_1000ev_lowfe}Distribution of the estimated fake yield in toy MC samples consisting of 1000 events with two loose leptons per event, for the standard matrix method, the standard matrix method with average efficiencies,  and the likelihood matrix method.  In these samples, the average value for $r$ is 0.90 and for $f$ is 0.20, with each efficiency having a Gaussian-distributed spread of 0.10.  The four plots show samples with different average values for the true number of events with fake leptons, as indicated by the dashed lines. }
\end{figure}

\begin{figure}
  \begin{center}
  \subfloat[]{\includegraphics[width=0.45\textwidth]{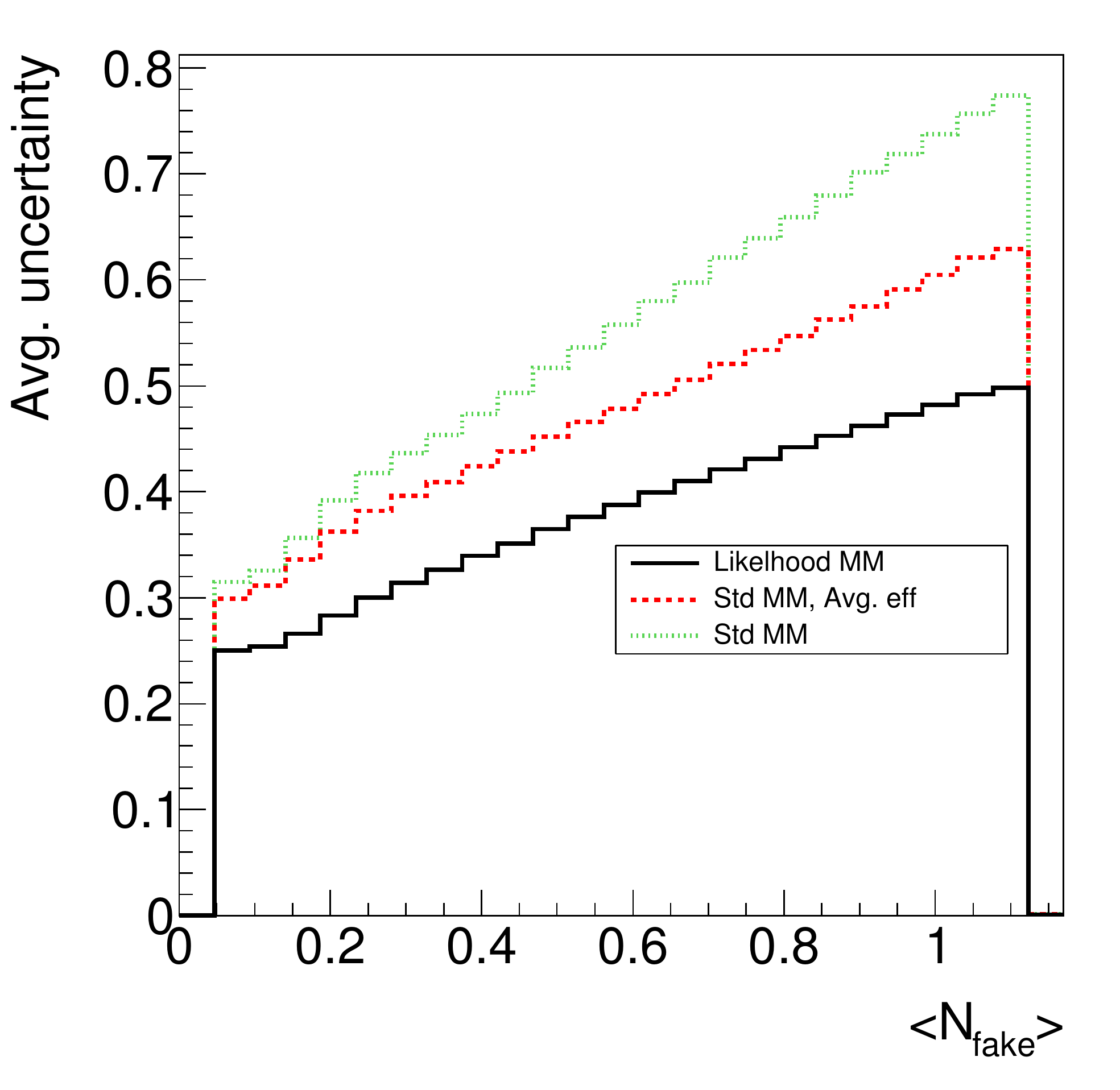}}
  \subfloat[]{\includegraphics[width=0.45\textwidth]{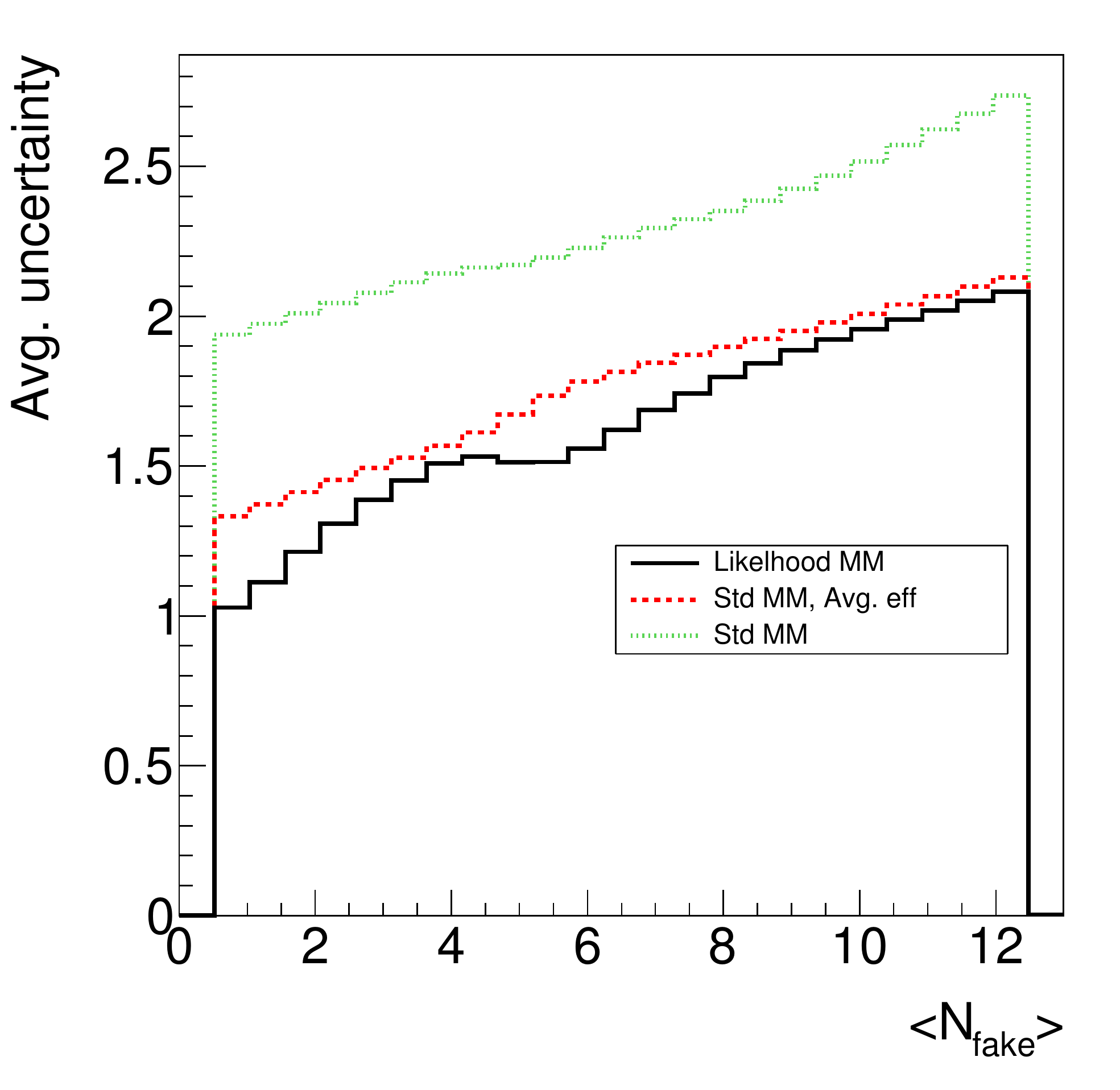}}\\
\subfloat[]{ \includegraphics[width=0.45\textwidth]{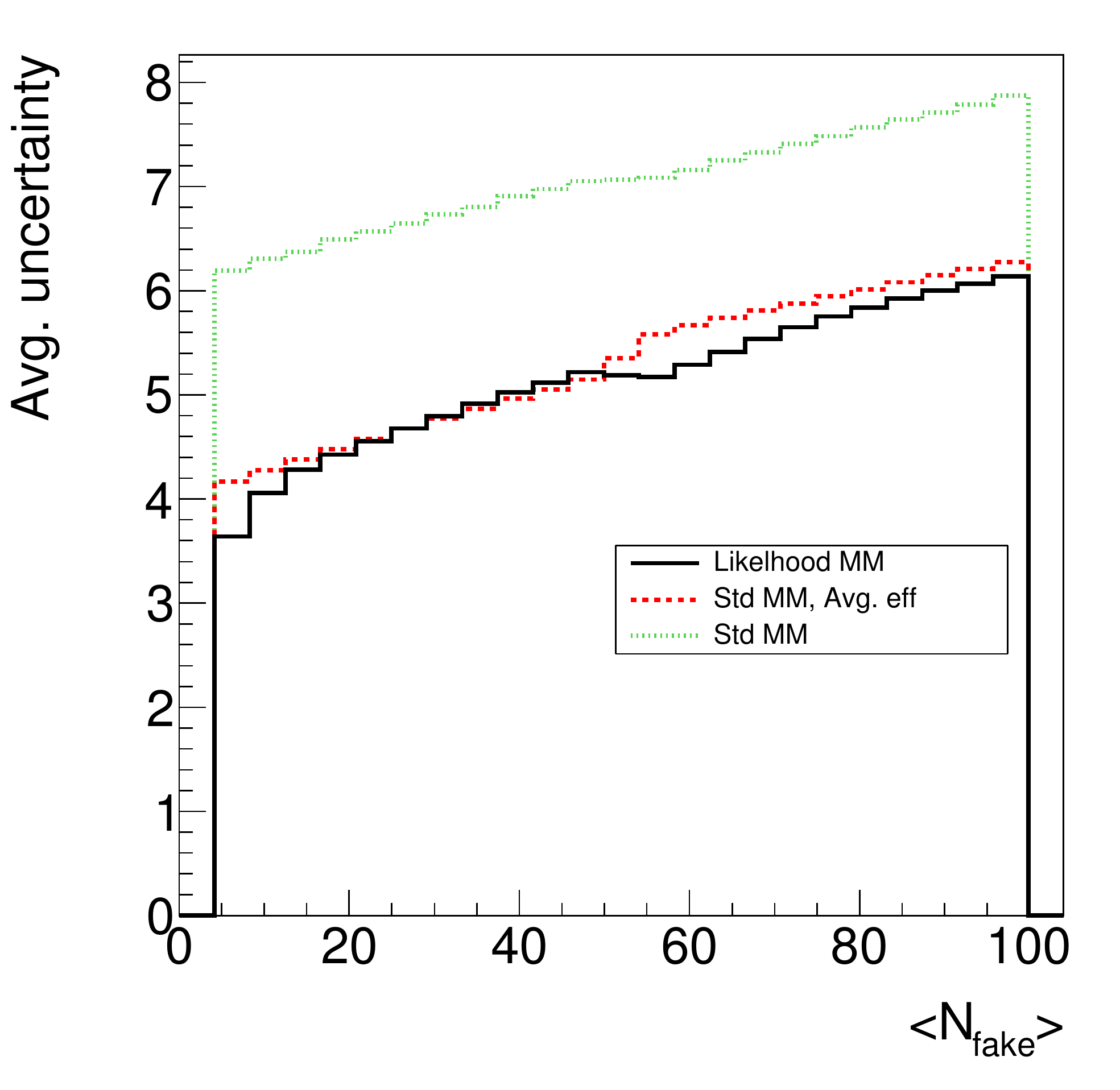}}
  \end{center}
  \caption{\label{fig:nfake_err_lowfe}Average uncertainties on $N^{tt}_{\mathrm {fake}}$ in simulated samples of dilepton events as a function of the average number of events with a fake lepton, for the standard matrix method, the standard matrix method with average efficiencies, and the likelihood matrix method.  Plot (a) shows pseudoexperiments with five loose dilepton events, plot (b) shows pseudoexperiments with 100 loose dilepton events, and plot (c) shows pseudoexperiments with 1000 loose dilepton events.   In these samples, the average value for $r$ is 0.90 and for $f$ is 0.20, with each efficiency having a Gaussian-distributed spread of 0.10.  }
\end{figure}

Figures~\ref{fig:nfake_5ev_lowfe}-\ref{fig:nfake_1000ev_lowfe} show the results for the standard matrix method, the standard matrix method with average efficiencies, and the likelihood approach, for simulated dilepton samples with 5, 100, and 1000 events and where the average values of $r$ and $f$ are 0.9 and 0.2, respectively.  The expected advantages of the likelihood approach are reflected in these plots: the likelihood approach never returns a negative value for the estimated fake lepton yield, and at for low-statistics samples the likelihood estimate is clustered much more closely around the true value.  The improvement in the precision of the likelihood approach persists to some extent even when statistics are large, as shown in Fig.~\ref{fig:nfake_1000ev_lowfe}. Figure~\ref{fig:nfake_err_lowfe} shows the uncertainty for both the standard and likelihood approaches as a function of the true number of fake leptons in each pseudoexperiment, for three different values of the number of events in each loose lepton sample (5, 100, and 1000).  In all cases the uncertainty from the likelihood approach is smaller than that from the standard approach.  This is likely due to the presence of terms of the form $r-f$ in the denominator of the event weights calculated in the standard approach; if $r$ and $f$ happen to have nearly the same values even for a small fraction of leptons in the sample, the events with these leptons are assigned large weight values, which can lead to instability in the result.

\begin{figure}
  \begin{center}
  \includegraphics[width=\textwidth]{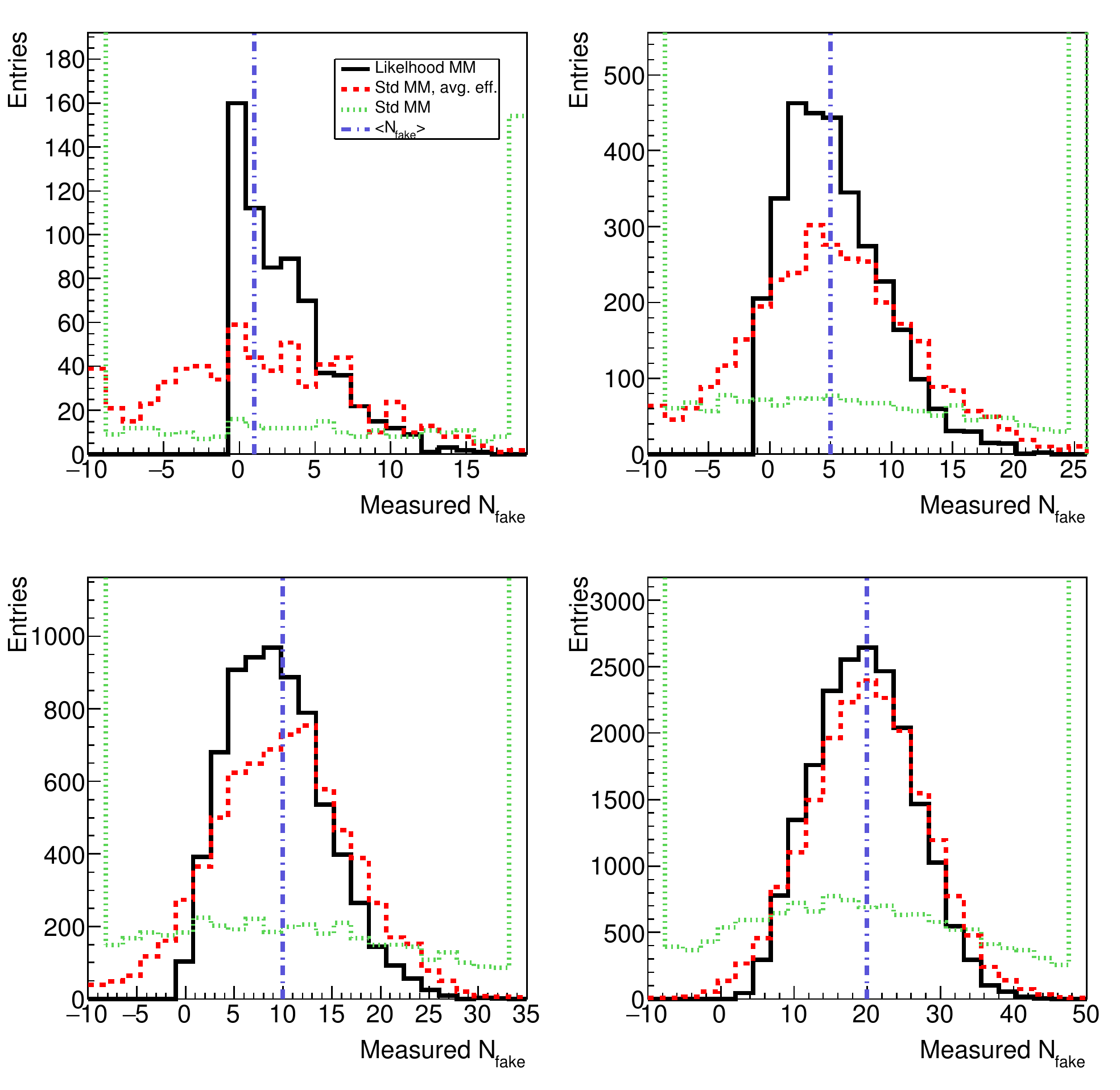}
  \end{center}
  \caption{\label{fig:nfake_100ev_medfe}Distribution of the estimated fake yield in toy MC samples consisting of 100 events with two loose leptons per event, for the standard matrix method, the standard matrix method with average efficiencies, and the likelihood matrix method.  In these samples, the average value for $r$ is 0.90 and for $f$ is 0.50, with each efficiency having a Gaussian-distributed spread of 0.10.  The four plots show samples with different average values for the true number of events with fake leptons, as indicated by the dashed lines. }
\end{figure}

\begin{figure}
  \begin{center}
  \includegraphics[width=\textwidth]{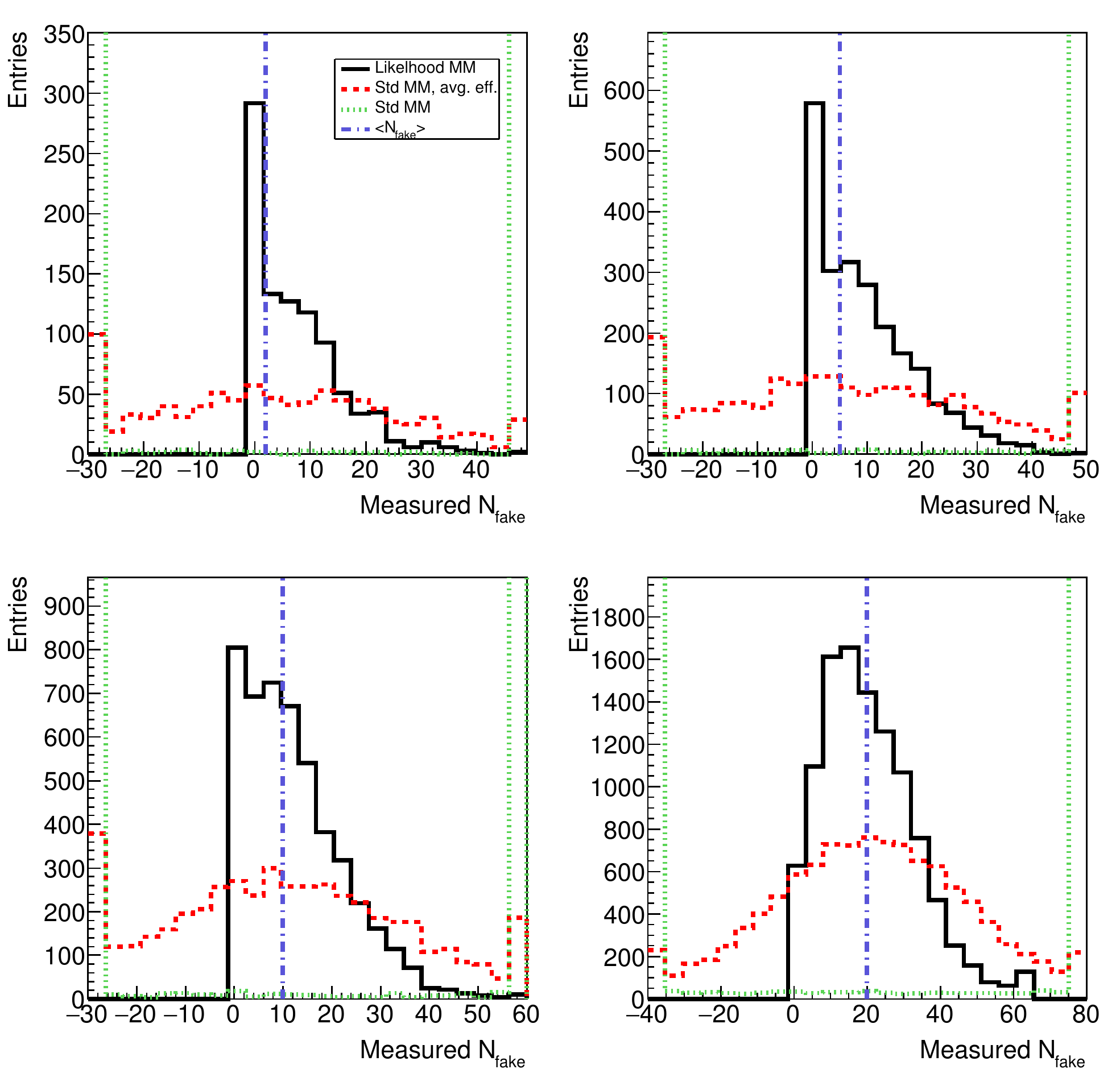}
  \end{center}
  \caption{\label{fig:nfake_100ev_hife}Distribution of the estimated fake yield in toy MC samples consisting of 100 events with two loose leptons per event, for the standard matrix method, the standard matrix method with average efficiencies,  and the likelihood matrix method.  In these samples, the average value for $r$ is 0.90 and for $f$ is 0.70, with each efficiency having a Gaussian-distributed spread of 0.10.  The four plots show samples with different average values for the true number of events with fake leptons, as indicated by the dashed lines. }
\end{figure}

As the luminosity of the LHC increases, it is likely that more restrictive trigger criteria will have to be applied to electron and muon candidates to keep the event rate within the available bandwidth.  As a result, the subset of fake leptons that satisfy the trigger requirements will be more signal-like, meaning the value of $f$ will tend to be larger.  Therefore it is worthwhile to explore the performance of both the likelihood and standard matrix method approaches for larger values of $f$.  Figures~\ref{fig:nfake_100ev_medfe} and~\ref{fig:nfake_100ev_hife} show the results for simulated 100-event dilepton samples with average values of $f$ of 0.5 and 0.7, respectively. The performance of the likelihood approach is degraded with respect to that shown in Fig.~\ref{fig:nfake_100ev_lowfe}, as expected since the ability to distinguish real from fake leptons has been diminished.  However, the performance remains reasonable, with the estimated fake lepton yield clustered around the true value.  In contrast, the numerical instability mentioned above for the standard likelihood approach is greatly exacerbated, since the fraction of lepton candidates for which $r \approx f$ is much greater.  This renders the standard approach essentially unusable, since the estimated fake yield is often driven to large positive or negative values.

\begin{figure}
\begin{center}
  \includegraphics[width=\textwidth]{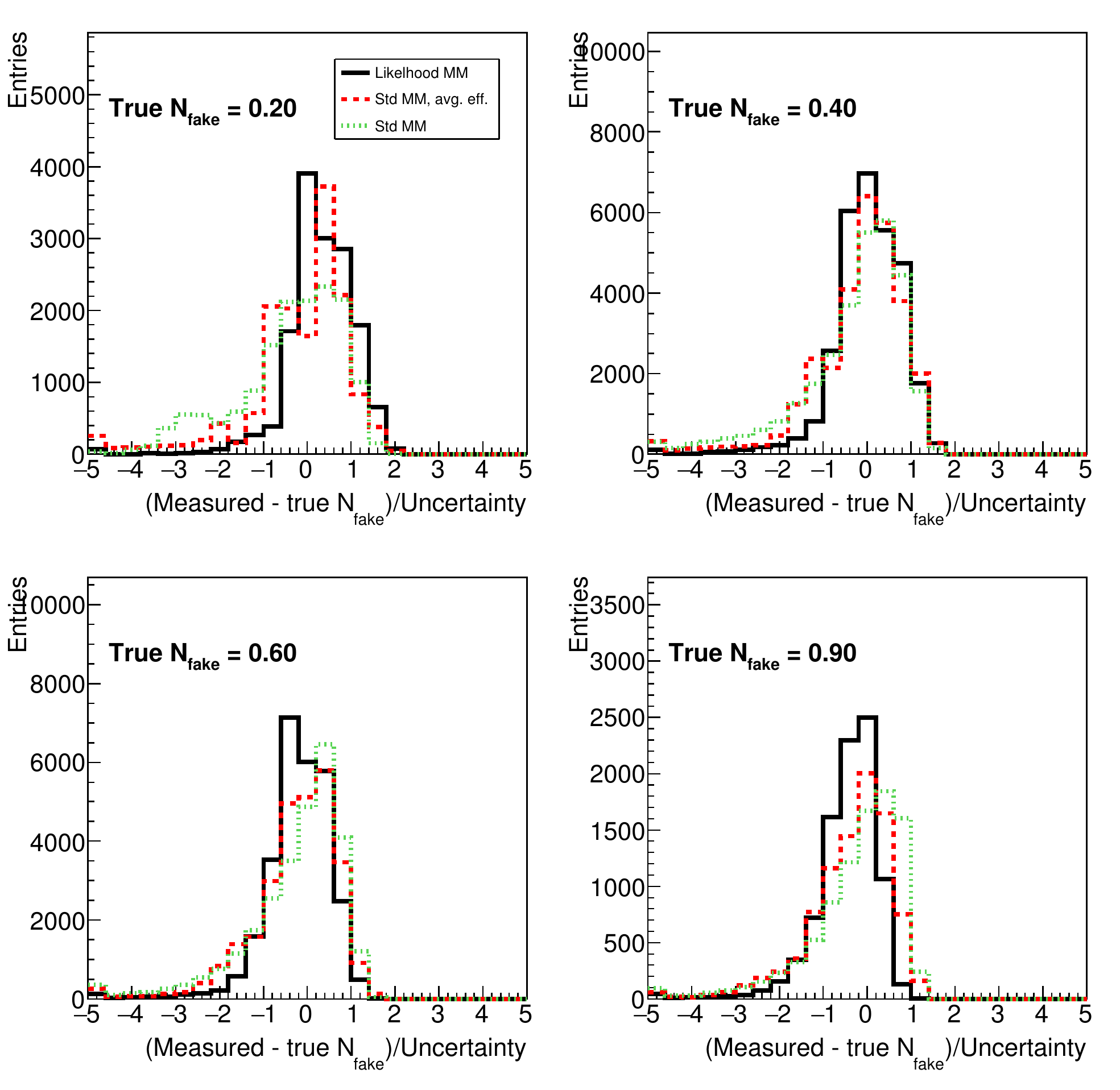}
\end{center}
  \caption{\label{fig:nfake_pull_5ev_lowfe}Pull distributions for the estimated fake yield in toy MC samples consisting of five events with two loose leptons per event, for the  standard matrix method, the standard matrix method with average efficiencies,  and the likelihood matrix method.  In these samples, the average value for $r$ is 0.90 and for $f$ is 0.20, with each efficiency having a Gaussian-distributed spread of 0.10.  The four plots show samples with different average values for the true number of events with fake leptons, as indicated on the plots.}
\end{figure}

\begin{figure}
  \begin{center}
  \includegraphics[width=\textwidth]{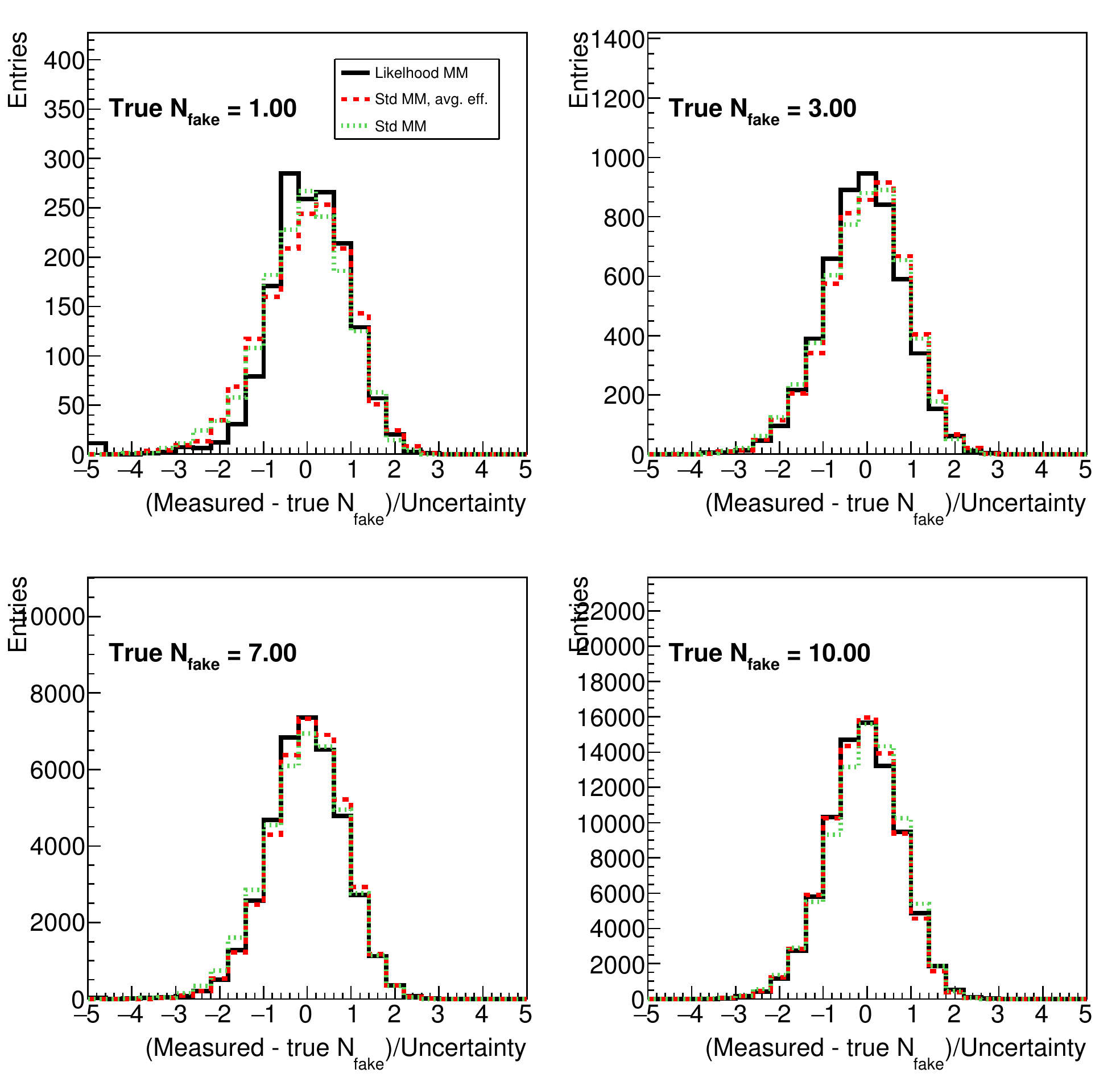}
  \end{center}
  \caption{\label{fig:nfake_pull_100ev_lowfe} Pull distributions for the estimated fake yield in toy MC samples consisting of 100 events with two loose leptons per event, for the standard matrix method, the standard matrix method with average efficiencies,  and the likelihood matrix method.  In these samples, the average value for $r$ is 0.90 and for $f$ is 0.20, with each efficiency having a Gaussian-distributed spread of 0.10.  The four plots show samples with different average values for the true number of events with fake leptons, as indicated on the plots.}
\end{figure}

\begin{figure}
  \begin{center}
  \includegraphics[width=\textwidth]{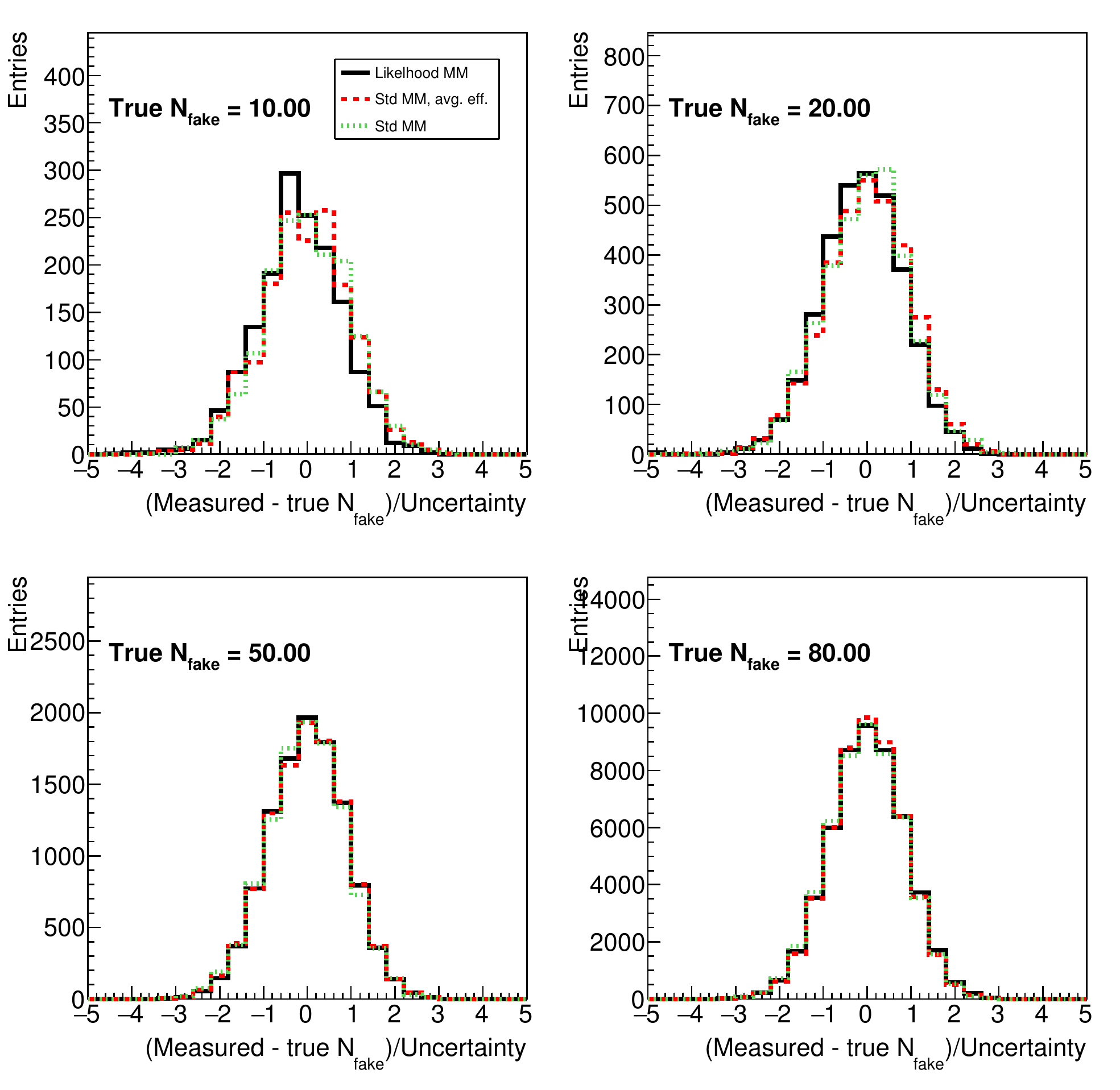}
  \end{center}
  \caption{\label{fig:nfake_pull_1000ev_lowfe}Pull distributions for the estimated fake yield in toy MC samples consisting of 1000 events with two loose leptons per event, for  the standard matrix method, the standard matrix method with average efficiencies,  and the likelihood matrix method.  In these samples, the average value for $r$ is 0.90 and for $f$ is 0.20, with each efficiency having a Gaussian-distributed spread of 0.10.  The four plots show samples with different average values for the true number of events with fake leptons, as indicated on the plots. }
\end{figure}

\begin{figure}
  \begin{center}
  \includegraphics[width=\textwidth]{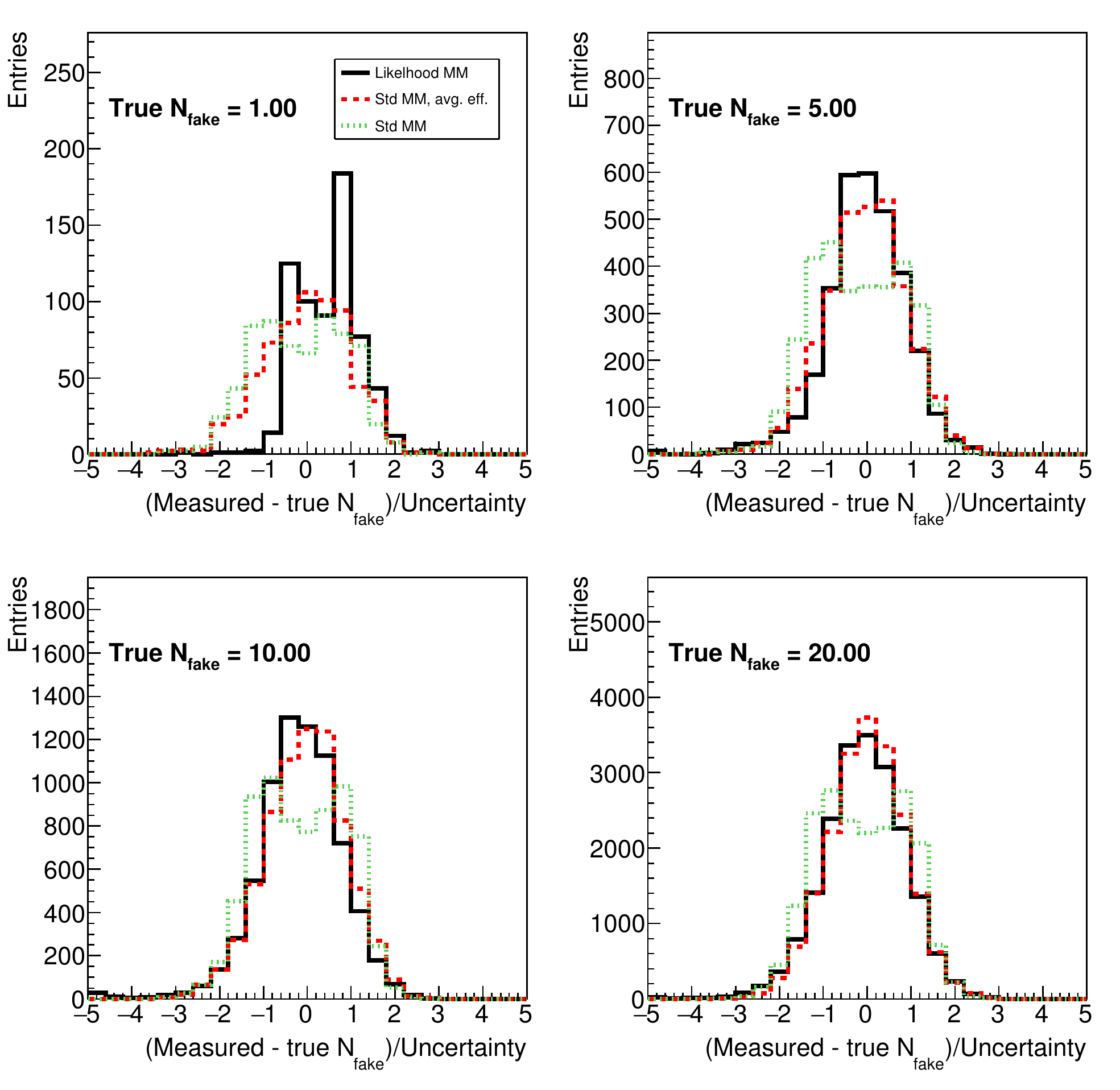}
  \end{center}
  \caption{\label{fig:nfake_pull_100ev_medfe}Pull distributions for the estimated fake yield in toy MC samples consisting of 100 events with two loose leptons per event, for the standard matrix method, the standard matrix method with average efficiencies,  and the likelihood matrix method.  In these samples, the average value for $r$ is 0.90 and for $f$ is 0.50, with each efficiency having a Gaussian-distributed spread of 0.10.  The four plots show samples with different average values for the true number of events with fake leptons, as indicated on the plots. }
\end{figure}

\begin{figure}
  \begin{center}
  \includegraphics[width=\textwidth]{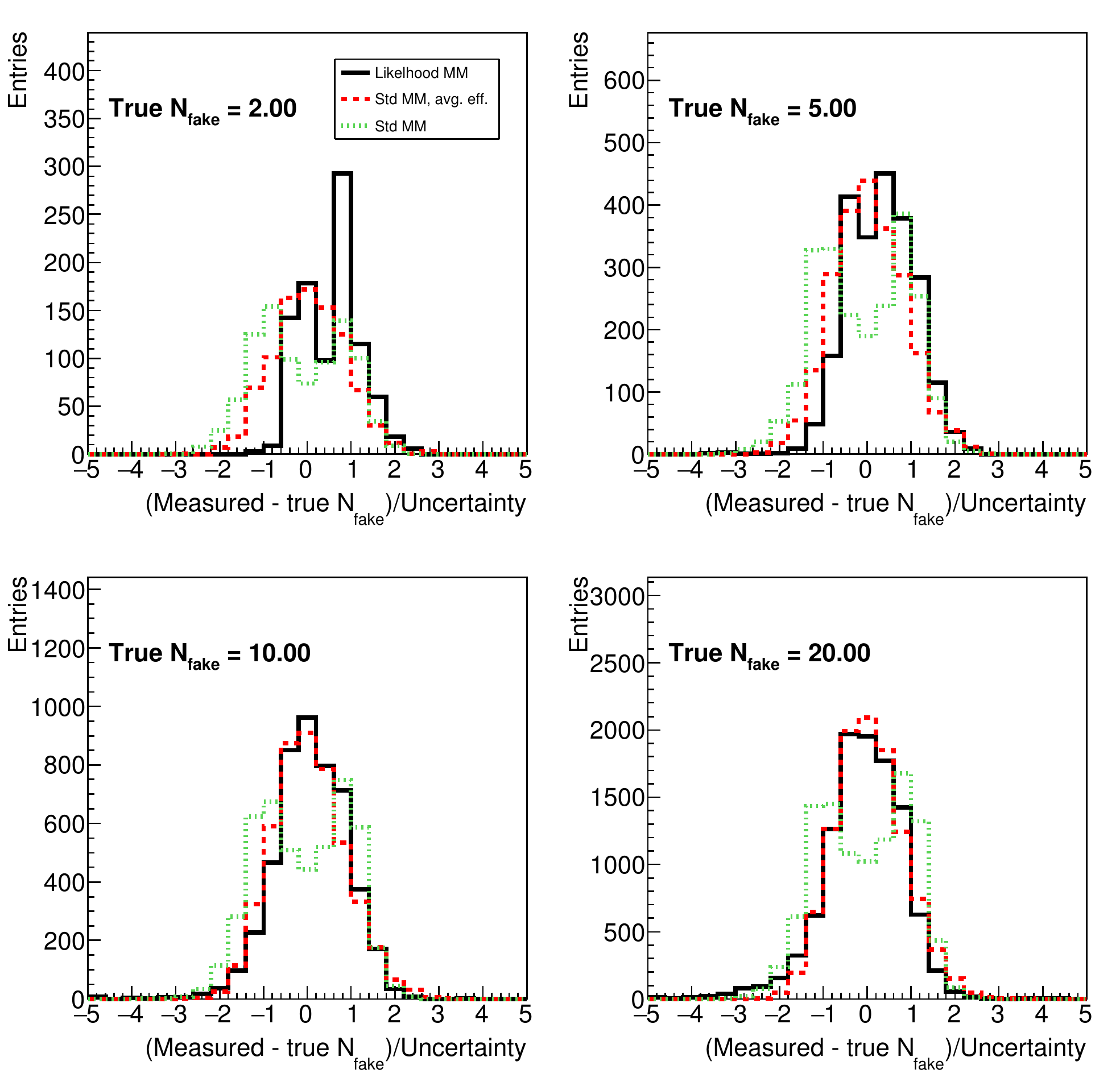}
  \end{center}
  \caption{\label{fig:nfake_pull_1000ev_hife}Pull distributions for the estimated fake yield in toy MC samples consisting of 1000 events with two loose leptons per event, for the standard matrix method, the standard matrix method with average efficiencies,  and the likelihood matrix method.  In these samples, the average value for $r$ is 0.90 and for $f$ is 0.70, with each efficiency having a Gaussian-distributed spread of 0.10.  The four plots show samples with different average values for the true number of events with fake leptons, as indicated on the plots. }
\end{figure}


\section{Results when more loose than tight leptons are permitted}
\label{sec:moreLooseThanTight}

\begin{figure}
  \begin{center}
\subfloat[]{  \includegraphics[width=0.49\textwidth]{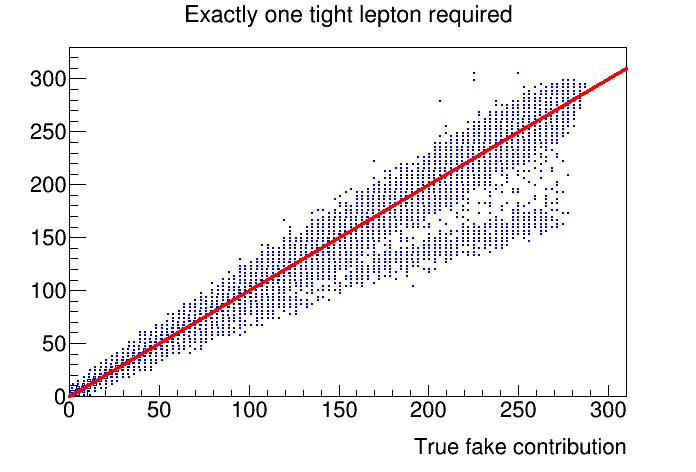}}
\subfloat[]{  \includegraphics[width=0.49\textwidth]{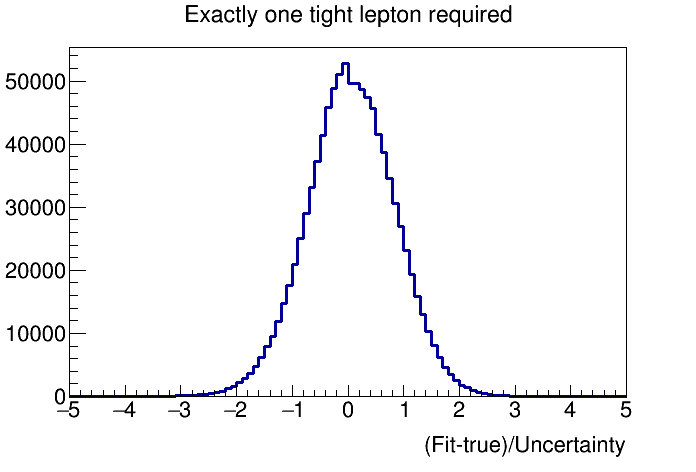}} \\
  \end{center}
  \caption{\label{fig:onelep_excl}Results of the fit when exactly one tight lepton is required, but up to three loose leptons are allowed.  The toy MC samples used assumed average real and fake lepton efficiencies of 0.9 and 0.2, respectively, and consisted of 1000 events per pseudoexperiment, with equal contributions from events with one, two, and three loose leptons. Plot (a) shows the output of the fit vs the true number of fakes in each pseudoexperiment, and plot (b) shows the pull distribution.}
\end{figure}

\begin{figure}
  \begin{center}
 \subfloat[]{\includegraphics[width=0.49\textwidth]{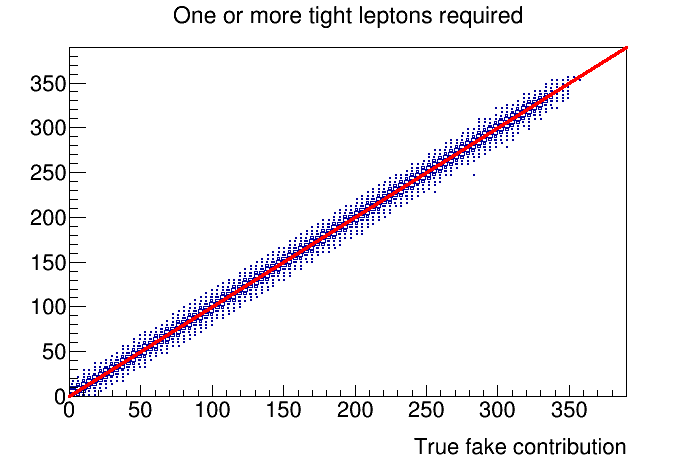}}
\subfloat[]{\includegraphics[width=0.49\textwidth]{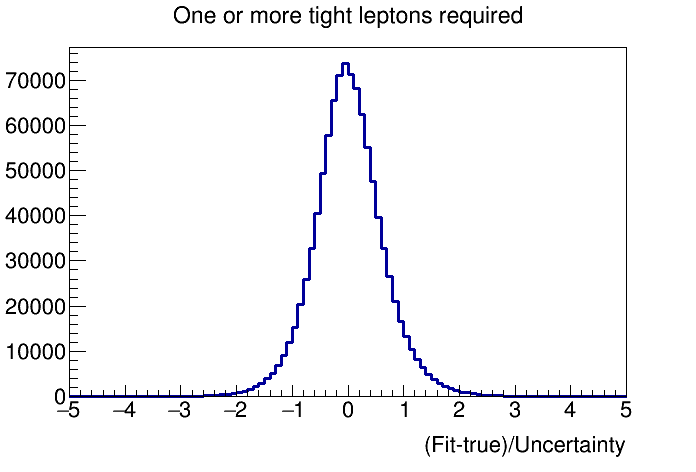}} \\
  \end{center}
  \caption{\label{fig:onelep_incl}Results of the fit when  one or more tight leptons are required, but up to three loose leptons are allowed.  The toy MC samples used assumed average real and fake lepton efficiencies of 0.9 and 0.2, respectively, and consisted of 1000 events per pseudoexperiment, with equal contributions from events with one, two, and three loose leptons. Plot (a) shows the output of the fit vs the true number of fakes in each pseudoexperiment, and plot (b) shows the pull distribution.}
\end{figure}

\begin{figure}
  \begin{center}
  \subfloat[]{ \includegraphics[width=0.49\textwidth]{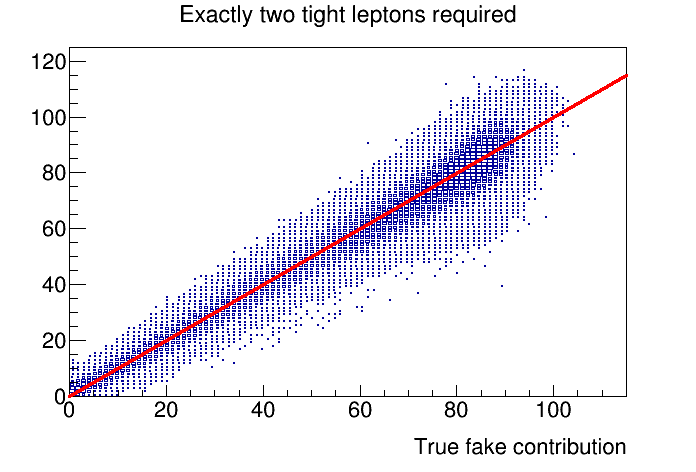}}
   \subfloat[]{\includegraphics[width=0.49\textwidth]{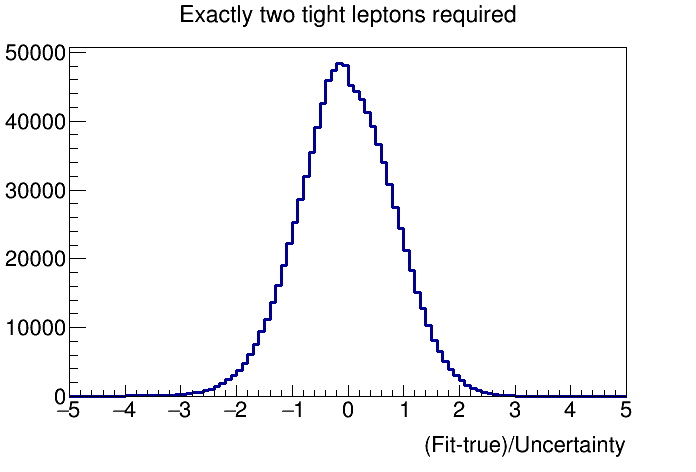}} \\
  \end{center}
  \caption{\label{fig:dilep_excl}Results of the fit when exactly two tight leptons are required, but up to three loose leptons are allowed.  The toy MC samples used assumed average real and fake lepton efficiencies of 0.9 and 0.2, respectively, and consisted of 1000 events per pseudoexperiment, with equal contributions from events with one, two, and three loose leptons. Plot (a) shows the output of the fit vs the true number of fakes in each pseudoexperiment, and plot (b) shows the pull distribution.}
\end{figure}

\begin{figure}
  \begin{center}
  \subfloat[]{ \includegraphics[width=0.49\textwidth]{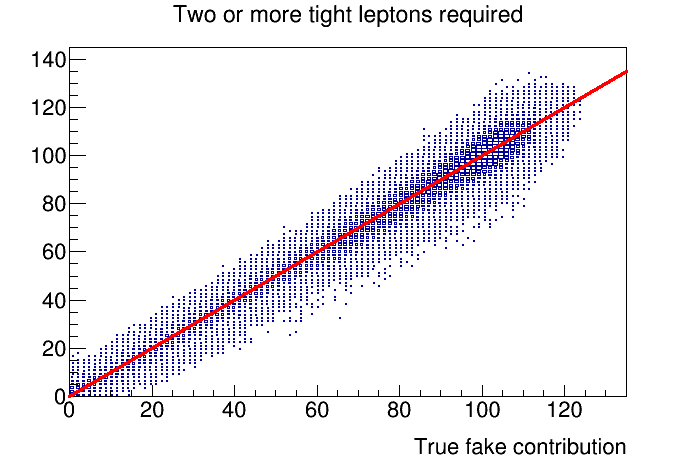}}
  \subfloat[]{ \includegraphics[width=0.49\textwidth]{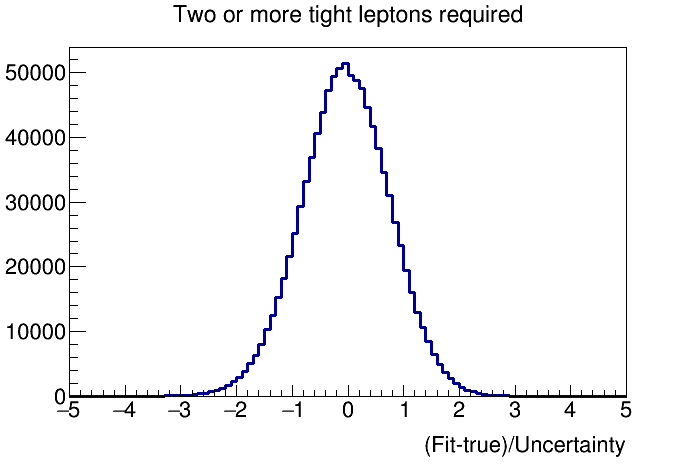}} \\
  \end{center}
  \caption{\label{fig:dilep_incl}Results of the fit when two or more tight leptons are required, but up to three loose leptons are allowed.  The toy MC samples used assumed average real and fake lepton efficiencies of 0.9 and 0.2, respectively, and consisted of 1000 events per pseudoexperiment, with equal contributions from events with one, two, and three loose leptons. Plot (a) shows the output of the fit vs the true number of fakes in each pseudoexperiment, and plot (b) shows the pull distribution.}
\end{figure}

Equations~\ref{eq:lhoodMM_nfake_params_1lepincl} and~\ref{eq:lhoodMM_nfake_params_1lepexcl} specify how the contribution of events with two loose leptons in analyses that require one tight lepton can be calculated.  The contributions from events with three loose leptons when one or two tight leptons are required can be computed similarly.  The total fake yield for an analysis that requires exactly one tight lepton is then the sum of the contributions from events with one, two, or three loose leptons~\footnote{And in principle, higher numbers of loose leptons, but cases beyond three loose leptons have not been implemented in the code.}  The performance of the method under all implemented combinations of the numbers of loose and tight leptons permitted is presented in Figs.~\ref{fig:onelep_excl}-~\ref{fig:dilep_incl}.
Figure~\ref{fig:onelep_excl} shows that the fit sometimes underestimates the true number of fake events if exactly one tight lepton is required but three loose leptons are allowed (this is the set of results that form a ``branch'' below the diagonal where most of the results are clustered).  About $10^{-3}$ of the fits fall into this category.  

\FloatBarrier

\section{Conclusions}

A Poisson likelihood approach for estimating the background yield due to fake leptons has been presented.  This method offers better precision and stability than the standard matrix method, especially in cases where there are substantial lepton-to-lepton variations in the real and fake lepton efficiencies or where the fake lepton efficiencies are large.  This method also has the advantage of ensuring a non-negative fake background yield estimate.  

\section{Acknowledgements}

I thank Simon Berlendis, Sarah Jones, Xiaowen Lei, Romain Madar, Fionnbarr O'Grady, and Zhaoxu Xi for helpful discussions and for testing the implementation of this method and reporting bugs in the software.

This work was supported by the Office of High Energy Physics of the U.S. Department of Energy through grant no. DE-SC0009913.

\bibliographystyle{elsarticle-num} 
\bibliography{LhoodMM.bib}

\end{document}